\documentclass[twocolumn,floatfix, showpacs]{revtex4}

\usepackage[final]{epsfig}
\usepackage{amsmath}
\usepackage{parskip}
 
\begin{document}
 
\title{Jet modification in 200 AGeV Au-Au collisions}
 
\author{Thorsten Renk}
\email{trenk@phys.jyu.fi}
%\author{Kari J.~Eskola}
%\email{kari.eskola@phys.jyu.fi}
\affiliation{Department of Physics, P.O. Box 35 FI-40014 University of Jyv\"askyl\"a, Finland}
\affiliation{Helsinki Institute of Physics, P.O. Box 64 FI-00014, University of Helsinki, Finland}
 
%\preprint{HIP-2006-46/TH}

\begin{abstract}
The computation of hard processes in hadronic collisions is a major success of perturbative Quantum Chromodynamics (pQCD). The environment of heavy-ion collisions offers the opportunity to embed such hard processes into a soft medium which is created simultaneously and study the medium-induced modifications. On the level of single high transverse momentum ($P_T$) hadrons, a reduction in yield, the so-called quenching is observed. However, on the level of jets, the energy-momentum flux carried by hadrons is conserved, i.e. the effect of the medium is a redistribution of energy and momentum, and statements about quenching of jets can only be made for specific cuts used to identify the jet. In this paper, we present a baseline computation for the expected modification of the longitudinal momentum distribution in jets by the medium created in 200 AGeV Au-Au collisions given a realistic set of experimental cuts used to identify jets in a heavy-iron environment. For this purpose, we use a Monte-Carlo code developed to simulate in-medium shower evolution following a hard process and a 3-d hydrodynamical evolution to simulate the soft medium.
\end{abstract}
 
\maketitle

\section{Introduction}

Jet quenching, i.e.\ the energy loss of hard partons created in the first moments of a heavy ion collision due to interactions with the surrounding soft medium  has long been regarded a promising tool to study properties of the soft medium \cite{Jet1,Jet2,Jet3,Jet4,Jet5,Jet6}. The basic idea is to study the changes induced by the medium to a hard process which is well-known from p-p collisions. A number of observables is available for this purpose, among them suppression in single inclusive hard hadron spectra $R_{AA}$ \cite{PHENIX_R_AA}, the suppression of back-to-back correlations \cite{Dijets1,Dijets2} or single hadron suppression as a function of the emission angle with the reaction plane \cite{PHENIX-RP}.

Single hadron observables and back-to-back correlations are well described in detailed model calculations using the concept of energy loss \cite{HydroJet1,Dihadron1,Dihadron2}, i.e. under the assumption that the process can be described by medium-induced radiation which shifts the leading parton energy by an energy $\Delta E$, followed by a fragmentation process using vacuum fragmentation with the shifted energy. However, there are also calculations for these observables in which the evolution of the in-medium parton distribution is followed in an analytic way \cite{HydroJet2,HydroJet3,Dihadron3}. Recently, also Monte Carlo (MC) codes for in-medium shower evolution have become available \cite{JEWEL,YAS}. In such computations, energy is not simply lost but redistributed in a characteristic way.

Jet observables (as opposed to single hadron measurements) allow in principle to probe this redistribution of energy through the interaction with the medium in detail. In this paper, we aim to make a benchmark calculation for the modification of the longitudinal momentum distribution of jets in 200 AGeV central Au-Au collisions at RHIC, based on the jet finding strategy used by the STAR collaboration \cite{STAR-Jets}. This observable in particular is expected to be sensitive to energy redistribution to increased hadron production at low $P_T$ due to medium-enhanced branchings \cite{HBP}.

\section{Modelling jets in medium}

We describe jet production in several steps. First we compute the production of hard partons in leading order (LO) pQCD. These primary partons subsequently evolve as a parton shower and we assume that it is this partonic evolution which is modified by the medium. Eventually the parton shower hadronizes, and we assume that this process takes place sufficiently far outside the medium to be treated as in vacuum. The hadronized parton shower constitutes a jet in our model and is analyzed using the experimental jet finding strategy. A detailed description of the model is found in \cite{YAS}, here we just summarize the main steps.

The production of two hard back to back partons $k,l$ with momentum $p_T$ in a p-p or A-A collision in LO pQCD is given by
 
\begin{equation}
\label{E-2Parton}
\frac{d\sigma^{AB\rightarrow kl +X}}{d p_T^2 dy_1 dy_2} \negthickspace = \sum_{ij} x_1 f_{i/A} 
(x_1, Q^2) x_2 f_{j/B} (x_2,Q^2) \frac{d\hat{\sigma}^{ij\rightarrow kl}}{d\hat{t}}
\end{equation}
 
where $A$ and $B$ stand for the colliding objects (protons or nuclei) and $y_{1(2)}$ is the 
rapidity of parton $k(l)$. The distribution function of a parton type $i$ in $A$ at a momentum 
fraction $x_1$ and a factorization scale $Q \sim p_T$ is $f_{i/A}(x_1, Q^2)$. The distribution 
functions are different for the free protons \cite{CTEQ1,CTEQ2} and protons in nuclei 
\cite{NPDF,EKS98}. The fractional momenta of the colliding partons $i$, $j$ are given by
%\begin{equation}
$ x_{1,2} = \frac{p_T}{\sqrt{s}} \left(\exp[\pm y_1] + \exp[\pm y_2] \right)$.
%\end{equation}
 
Expressions for the pQCD subprocesses $\frac{d\hat{\sigma}^{ij\rightarrow kl}}{d\hat{t}}(\hat{s}, 
\hat{t},\hat{u})$ as a function of the parton Mandelstam variables $\hat{s}, \hat{t}$ and $\hat{u}$ 
can be found e.g. in \cite{pQCD-Xsec}.
To account for various effects, including higher order pQCD radiation, transverse motion of partons in the nucleon (nuclear) wave function and effectively also the fact that hadronization is not a collinear process, we fold into the distribution an intrinsic transverse momentum $k_T$ with a Gaussian distribution of width 1.6 GeV. This momentum vector points into a random direction in the transverse plane and modifies the transverse momenta $p_{T_1}, p_{T_2}$ of the outgoing partons in creating a momentum imbalance between them according to $p_{T_1} + p_{T_2} =  k_T$.

The probability density $P(x_0, y_0)$ for finding a hard vertex  in an A-A collision at the 
transverse position ${\bf r_0} = (x_0,y_0)$ and impact 
parameter ${\bf b}$ is given by the product of the nuclear profile functions as
\begin{equation}
\label{E-Profile}
P(x_0,y_0) = \frac{T_{A}({\bf r_0 + b/2}) T_A(\bf r_0 - b/2)}{T_{AA}({\bf b})},
\end{equation}
where the thickness function is given in terms of Woods-Saxon the nuclear density
$\rho_{A}({\bf r},z)$ as $T_{A}({\bf r})=\int dz \rho_{A}({\bf r},z)$. For the present study, we evaluate Eq.~(\ref{E-Profile}) at ${\bf b} = 0$ corresponding to central collisions.

Inclusive production of a parton flavour $f$ at rapidity 
$y_f$ and momentum $p_T$ is found from Eq.~(\ref{E-2Parton}) by integrating over either $y_1$ or $y_2$ and summing over appropriate combinations of partons. We use this distribution probed at midrapidity as an input for the next step, the evolution of a shower.

In vacuum, we simulate using the PYSHOW algorithm \cite{PYSHOW} which is part of PYTHIA \cite{PYTHIA}. To take into account medium effects, we assume a medium which does not absorb momentum by recoil of its constituents, but rather increases the virtuality of partons propagating through it and thus inducing additional radiation. Such a medium can be characterized by a transport coefficient $\hat{q}$ which represents the increase in virtuality $\Delta Q^2$ per unit pathlength of a parton traversing the medium. 

Since the pQCD shower evolution takes place in momentum space, in order to incorporate this type of medium modification, we need to make a link to the spacetime evolution of the shower. 
We assume that the formation time of a shower parton with virtuality $Q$ is developed on the timescale $1/Q$, i.e. the lifetime of a virtual parton with virtuality $Q_b$ coming from a parent parton with virtuality $Q_a$ is in the rest frame of the original hard collision (the rest frame of the medium may be different by a flow boost as the medium may not be static) given by
$\tau_b = \frac{E_b}{Q_b^2} - \frac{E_b}{Q_a^2}$.

The time $\tau_a^0$ at which a parton $a$ is produced in a branching can be determine by summing the lifetimes of all ancestors. Thus, during its lifetime, the parton virtuality is increased by the amount
\begin{equation}
\label{E-Qgain}
\Delta Q_a^2 = \int_{\tau_a^0}^{\tau_a^0 + \tau_a} d\zeta \hat{q}(\zeta)
\end{equation}
where $\zeta$ is integrated along the spacetime path of the parton through the medium and $\hat{q}(\zeta) = \hat{q}(\tau,r,\phi,\eta_s)$ describes the spacetime variation of the transport coefficient where $\hat{q}$ is specified at each set of coordinates proper time $\tau$, radius $r$, angle $\phi$ and spacetime rapidity $\eta_s$. In the following, we will use a hydrodynamical evolution model of the medium \cite{Hydro3d} for this dependence assuming that $\hat{q}$ scales as
\begin{equation}
\label{E-qhat}
\hat{q}(\xi) = K \cdot 2 \cdot \epsilon^{3/4}(\zeta) (\cosh \rho - \sinh \rho \cos\alpha)
\end{equation}
with the energy density $\epsilon$ and the local flow rapidity $\rho$ with angle $\alpha$ between flow and parton trajectory.
If the parton is a gluon, the virtuality transfer from the medium is increased by the ratio of gluon to quark Casimir color factors, $3/\frac{4}{3} =  2.25$.

The additional virtuality transfer to the parton in Eq.~(\ref{E-Qgain}) leads to an increased branching rate and thus to both to an increase of the angular spread of the shower with respect to the original hard parton trajectory and a reduction in hard parton production due to increased production of soft partons \cite{YAS}.
Note that Eq.~(\ref{E-Qgain})  needs to evaluated for every parton path through the medium from the hard vertex, Eq.~(\ref{E-Profile}). However, in \cite{YAS} it was found that to good approximation it is sufficient to sort paths according to $\Delta Q^2_{tot}$, the integrated virtuality along the eikonal path of the original hard parton through the medium. This simplifies the computation considerably.

In a last step, the Lund string fragmentation scheme \cite{Lund} is used to hadronize the shower. Note that, as discussed in \cite{YAS}, while the assumption that hadronization can be assumed to be unmodified by the medium is expected to hold for high $P_T$ light hadrons, it may break down for heavy hadrons produced at low momentum, thus this end of the distribution is computed less reliably.

\section{Jet identification and medium-induced modification}

\begin{figure*}[htb]
\epsfig{file=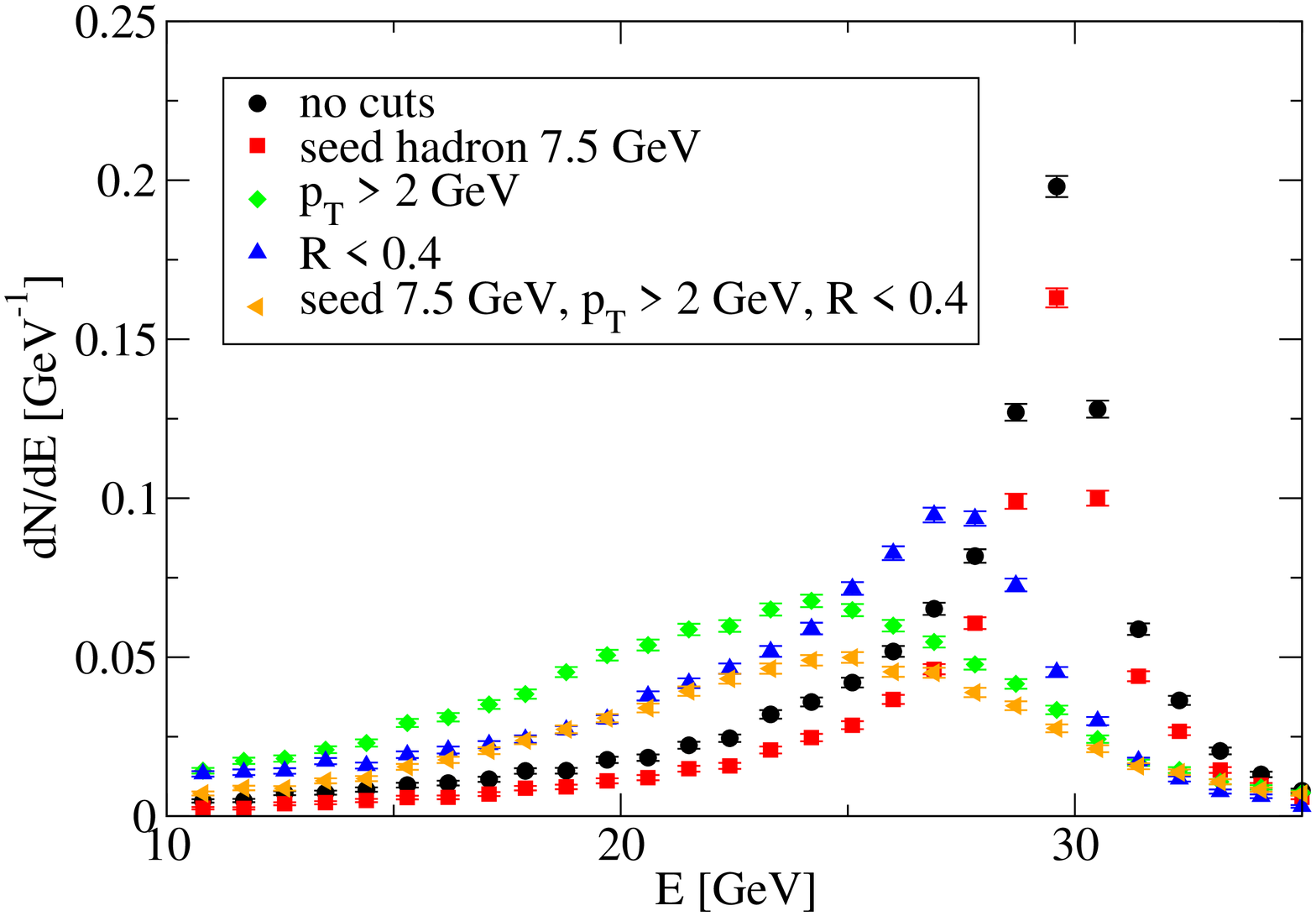, width=8cm}\epsfig{file=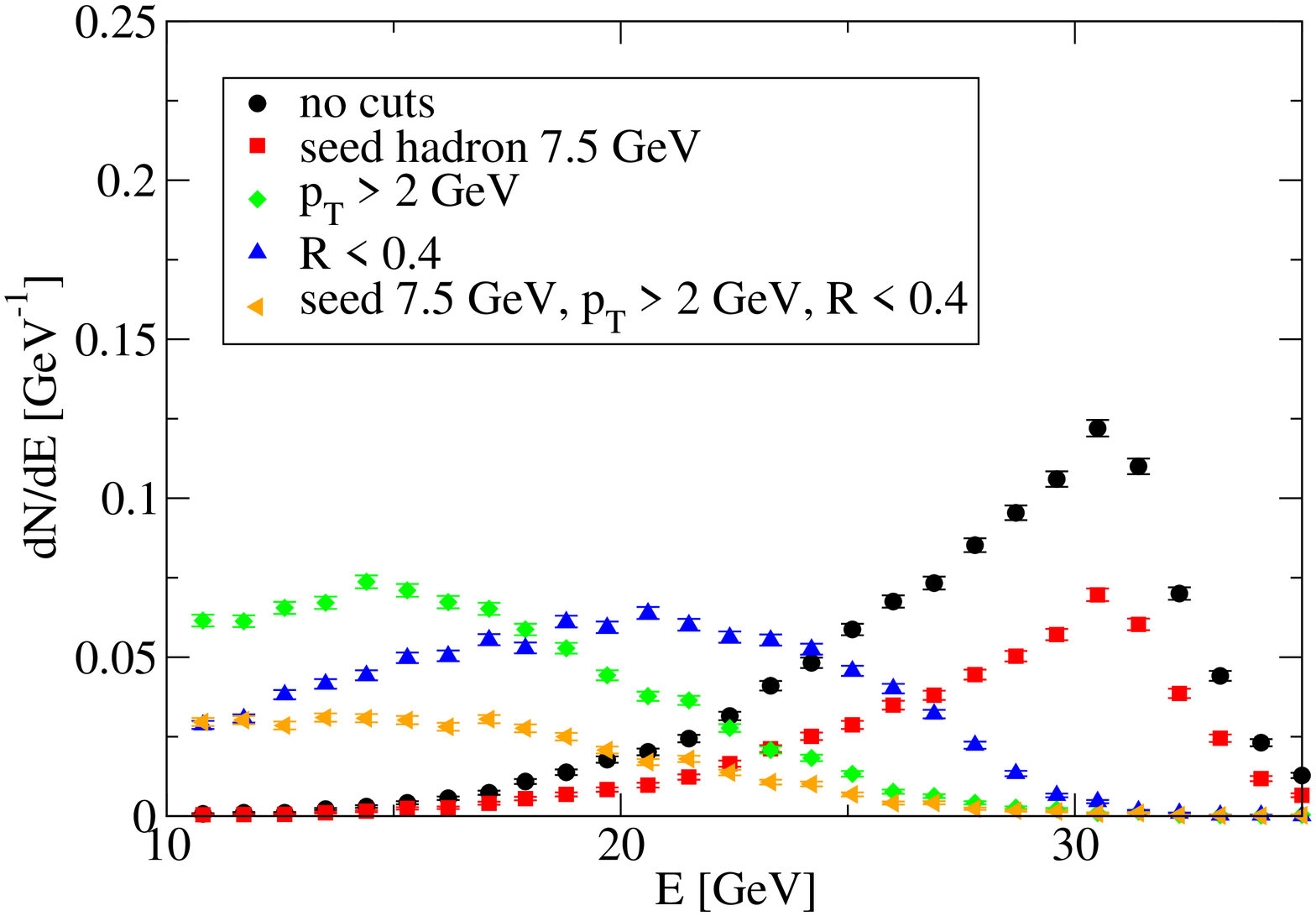, width=8cm}
\caption{\label{F-dNdE}(Color online) Distribution of the energy with which a quark jet of 30 GeV energy is detected $dN/dE$, given different cuts in hadron momenta or cone width. Left panel: without medium modification. Right panel: medium modified with $\Delta Q^2_{tot}=15$ GeV$^2$ (see text), roughly corresponding to a path through the whole medium.}
\end{figure*}

Experimentally, jets are identified using calorimetric measurements in addition to a series of cuts to eliminate background contributions. Thus, for a given parton energy $E_p$ in the simulation, a detector does not always find a jet with the same energy. This introduces a bias on any properties of identified jets as compared to all jets. In the following, we study this problem simulating as closely as possible the cuts used by the STAR collaboration to find jets \cite{STAR-Jets}, although we stress that the formalism is completely general and could be used for any other set of cuts. In this section, we focus on $dN/dE_{E_p}$, the distribution of the detected jet energy $E$, given an underlying parton energy $E_p$.

For an ideal detector, one would expect this distribution to be $\delta(E-E_p)$ in the absence of intrinsic $k_T$ and a Gaussian centered around $E_p$ when taking into account inrinsic $k_T$. However, the STAR calorimetric measurement is only sensitive to charged hadrons, $\pi_0$ and $\gamma$, i.e. whenever the hadronization yields a different neutral hadron, its energy does not contribute to the reconstructed jet energy. This induces a long tail towards low energies.

Furthermore, in order to separate jets from the substantial background in a heavy-ion environment, the following additional requirements are imposed: (1) a seed hadron of 7.5 GeV or more is required to be in the shower as starting point for the jet finding algorithm (2) only hadrons with $P_T >2$ GeV are taken into account and (3) an angular cut with angle with respect to the jet axis of $R < 0.4$ is imposed. The effect of these cuts on $dN/dE_{E_p}$ is shown in Fig.~\ref{F-dNdE} for a quark jet in vacuum (left panel) and for a jet with $\Delta Q^2_{tot} = 15$ GeV$^2$ (right panel, corresponding to a path traversing the whole medium), all for $E_p=30$ GeV.
It is clearly seen how the medium acts to reduce the energy of the jet seen by the detector by reducing the probability that a seed hadron is found, by transporting energy out of the cone region $R<0.4$ and by increasing hadron production below the $P_T$ cut.

If we wish to study the medium suppression of jets observed in e.g. 25-30 GeV or 30-35 GeV momentum bins, we have to average the results of $dN/dE_{E_p}$ over different $E_p$ and parton type with the weight factors determined by Eq.~(\ref{E-2Parton}) and in addition average over all different trajectories from the vertex of origin Eq.~(\ref{E-Profile}) through the medium and then study which fraction of jets is found in a momentum bin, given the jet finding cuts, as compared to the situation in vacuum. In this way, a jet suppression factor $R_{AA}^{jet}$ can be defined. For this particular model of energy redistribution and the imposed cuts, $R_{AA}^{jet}$ is $0.43\pm0.04$ for the 25-30 GeV momentum bin and $0.46\pm 0.05$ for the 30-35 GeV bin (statistical errors only).

In order to study the change in the momentum distribution along the jet axis, we focus on $dN/d\xi$ where $\xi= \ln E/P_T$ is the logarithm of the jet energy divided by the energy of a hadron in the jet. When comparing this quantity for a single parton with known energy with and without the medium, a depletion at low $\xi$ corresponding to the reduction of high $P_T$ hadrons and an enhancement at high $\xi$ corresponding to low $P_T$ hadron production is expected \cite{JEWEL,YAS,HBP}.

However, the fact that a jet needs to pass the cuts to be identified experimentally introduces a strong bias on $dN/d\xi$. Events with multiple hadron production below the $P_T$ cut are strongly suppressed as compared to events with multiple hadron production just above the cut. Likewise, copious low $P_T$ hadron production reduces the chance to find a hard seed hadron.

\begin{figure}[hbt]
\epsfig{file=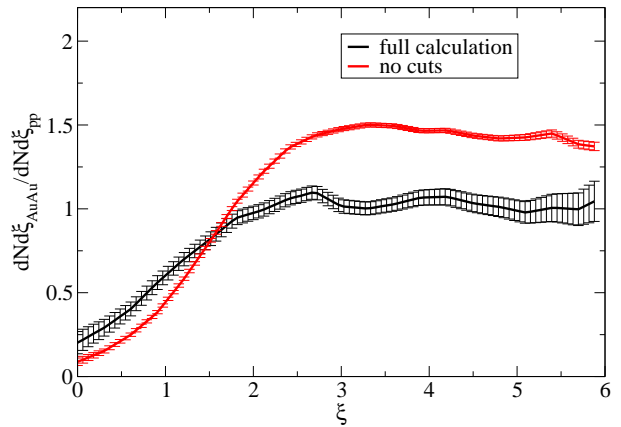,width=8cm}
\caption{\label{F-dNdxi}(Color online)The ratio of $dN/d\xi$ in central 200 AGeV Au-Au collisions over the vacuum value calculated for two different situations: Black only for jets which are found with an energy of 25-30 GeV within a cone of $R<0.4$ and for hadrons with $P_T>2$ GeV, but where the distribution was extracted without cut in $P_T$ and for a wider cone $R<0.7$. Red for fixed parton energy of 30 GeV assuming that all jets are detected without cuts.}
\end{figure}

In Fig.~\ref{F-dNdxi}, we present two different scenarios to illustrate this effect. First, we make a calculation close to the experimental situation in which the jet needs to pass the cuts discussed in the previous section and in which then, again in analogy with the experimental procedure, $dN/d\xi$ is extracted  for these jets for all $P_T$ in an increased cone with $R<0.7$. In the second scenario, we assume an ideal detector with no cuts imposed which allows to identify each produced jet regardless of its modification.

In both scenarios, we average over all possible paths through the medium and compute the ratio of the $dN/d\xi$ in medium and $dN/d\xi$ in vacuum. It is clearly seen how the trigger bias changes the expectations for larger $\xi$. Effectively, it prevents the observation of any enhancement in the region $\xi > 2$ where the unbiased calculation shows a strong effect.

On the other hand, the two scenarios are fairly similar towards $\xi=0$. In this region, the jet is effectively a single hadron carrying all the parton momentum. Such a jet is not influenced by any of the cuts and hence not biased. The fact that the value reached in the ratio is close to $R_{AA}$ for single hadron spectra is not an accident, as in this limit the suppression of single hard hadrons is probed.

\section{Discussion}

We have presented a benchmark calculation for the modification of jet properties, here the longitudinal momentum distribution of hadrons inside the jet, due to a medium as created in heavy-ion collisions. We find in principle sizeable modifications, characteristic for the pattern of energy redistribution in the medium, i.e. a strong depletion of high $P_T$ hadrons in the jet and an enhancement of low $P_T$ hadron production. However, in a more realistic calculation close to the experimental jet finding strategy, a strong trigger bias prevents one from observing the enhancement. The depletion of high $P_T$ hadrons in the jet however should still be visible (and agree with the single high $P_T$ hadron suppression).

While details of the calculation depend on the model used to simulate the medium effect, it should be stressed that the analysis framework is rather general and could be applied to any other model of in-medium jet modification (such as e.g. JEWEL \cite{JEWEL}). Likewise, the change of event structure induced by the trigger bias appears to be generic rather than a peculiarity of the model used here. Finally, a comparison with data would clearly be valuable to study in more detail at which point model assumptions like hadronization outside the medium even for low $P_T$ hadrons break down.

\begin{acknowledgments}
I'd like to thank J.~Putschke and P.~Jacobs for valuable discussions on the problem. This work was financially supported by the Academy of Finland, Project 115262. 
\end{acknowledgments}

\end{document}